\documentclass[useAMS,usenatbib]{mn2e}
\def\simless{\mathbin{\lower 3pt\hbox
{$\rlap{\raise 5pt\hbox{$\char'074$}}\mathchar"7218$}}}   
\def\simmore{\mathbin{\lower 3pt\hbox
{$\rlap{\raise 5pt\hbox{$\char'076$}}\mathchar"7218$}}}   

\def\Msun{M_\odot}                                       

\usepackage{afterpage}
\usepackage{tensor}
\usepackage{graphicx}
\usepackage{subfigure}
\usepackage{amsmath, amssymb, amsfonts}
\usepackage[usenames,dvipsnames]{color}
\usepackage{ulem}

\newcommand \risco {r_{\rm isco}}
\newcommand \tacc {t_{\rm acc}}

\newcommand \rg {r_{\rm g}}
\newcommand \rgc {\, r_{\rm g}/c}
\newcommand \rh {r_{\rm H}}
\newcommand \rclose {r_{\rm close}}
\newcommand \Pj {L_{\rm j}}
\newcommand \Lj {L_{\rm j}}
\newcommand \lcrit {l_{\rm crit}}
\newcommand \vk {v_{\rm K}}
\newcommand \vr {v_{r}}
\newcommand \cs {c_{\rm s}}

\newcommand \beq {\begin{equation}}
\newcommand \eeq {\end{equation}}
\newcommand \be {\begin{equation}}
\newcommand \ee {\end{equation}}

\newcommand \lp {\left(}
\newcommand \rp {\right)}

\newcommand \KP[1]{\textcolor{black}{#1}}

\let\oldhat\hat

\renewcommand{\hat}[1]{\oldhat{\mathbf{#1}}}

\newcommand\apjl{ApJ}
\newcommand\apj{ApJ}

\newcommand\aap{A\&A}
\newcommand\mnras{MNRAS}

\newcommand\nat{Nature}

\title[Black-hole jets without large-scale net magnetic flux]{Black-hole jets without large-scale net magnetic flux}

\author[Kyle Parfrey, Dimitrios Giannios, and Andrei M. Beloborodov]
{Kyle Parfrey$^{1}$\thanks{E-mail: parfrey@astro.princeton.edu}, Dimitrios Giannios$^2$, and  Andrei M. Beloborodov$^3$\\
$^{1}$Department of Astrophysical Sciences, Princeton University, Peyton Hall, Princeton, NJ 08544, USA\\
$^{2}$Department of Physics and Astronomy, Purdue University, 525 Northwestern
Avenue, West Lafayette, IN 47907, USA\\
$^{3}$Department of Physics and Columbia Astrophysics Laboratory, Columbia University, 538 West 120th Street, New York, NY 10027, USA}

\begin{document}
\maketitle

\label{firstpage}

\begin{abstract}
We propose a scenario for launching relativistic jets from rotating black holes, in which small-scale magnetic flux loops, sustained by disc turbulence, are forced to inflate and open by differential rotation between the black hole and the accretion flow. This mechanism does not require a large-scale net magnetic flux in the accreting plasma. Estimates suggest that the process could operate effectively in many systems, and particularly naturally and efficiently when the accretion flow is retrograde. We present the results of general-relativistic force-free electrodynamic simulations demonstrating the time evolution of the black hole's magnetosphere, the cyclic formation of jets, and the effect of magnetic reconnection. The jets are highly variable on timescales \KP{$\sim 10$--$10^3\,\rg/c$, where $\rg$ is the black hole's gravitational radius.} The reconnecting current sheets observed in the simulations may be responsible for the hard X-ray emission from accreting black holes.
\end{abstract} 
  
\begin{keywords}
black hole physics---galaxies: active---galaxies: jets---X-rays: binaries---gamma-ray burst: general---MHD
\end{keywords}

\section{Introduction} 
\label{intro}

Relativistic jets are ubiquitous throughout high-energy astrophysics. It is often assumed that large-scale magnetic fields
are responsible for jet launching, either from the inner region of an
accretion disc \citep{1982MNRAS.199..883B} or from a rotating black hole itself 
\citep{1977MNRAS.179..433B}. In the Blandford-Znajek (BZ) process, on which we focus in this Letter, the total jet power is \citep*{1982MNRAS.198..345M, 1986bhmp.book.....T}
\beq
\Pj \approx \Upsilon \lp\frac{\Phi a}{\rh}\rp^2 c \approx 4 \pi^2 \Upsilon \lp B_{\perp} \rh a\rp^2 c,
\label{eq:Pj}
\eeq
where $\Phi = (1/2) \int_{\theta\phi} |B^r| {\rm d}S_r \approx 2 \pi\rh^2 B_{\perp}$ is half the absolute magnetic flux on the black hole, whose mass, dimensionless spin parameter, horizon radius, and average horizon-normal magnetic field are $M$, $a$, $\rh = \lp1 + \sqrt{1-a^2}\rp \rg$, and $B_\perp$ respectively; $\rg = GM/c^2$. 
The dimensionless number $\Upsilon \approx 10^{-3}$ is determined by the magnetic field distribution on the horizon \citep[e.g.,][]{2010ApJ...711...50T}.

As typically applied, the BZ process requires (i) 
that the plasma supplied to the system be threaded by a net large-scale magnetic flux
sufficient to power the observed jet, as determined by equation~(\ref{eq:Pj}), and (ii) that this magnetic flux 
be advected to the immediate vicinity of the black hole. 
Neither of these conditions is trivially satisfied. By `large-scale' we mean that the characteristic size of the relevant flux system is already much greater than the black hole's gravitational radius before the two begin to interact. An extensive effort to simulate black-hole accretion discs and jets \citep[e.g.,][]{Koide:2000aa,2004ApJ...611..977M, 2006ApJ...641..103H} has generally found that large-scale fields are required or preferred for jet production \citep*{2008ApJ...678.1180B}.

In those systems where an estimate of the available magnetic flux can be made, it appears that
the field is insufficient to power the observed jets.
For example, long-duration gamma-ray bursts (GRBs) are associated with 
the core collapse of massive stars \citep{1998Natur.395..670G} and may be powered by $M\sim 3\,\Msun$ black holes. 
With typical jet power $\Pj \sim 10^{51}$~erg s$^{-1}$ and assuming a rapidly spinning black hole $a \sim 1$, the required magnetic flux is (see equation~\ref{eq:Pj}) $\Phi \sim 3\times 10^{27} L_{\rm{j},51}^{1/2}$~G~cm$^2$, corresponding to a field strength at the black hole of $B \sim 2\times 10^{15}$~G \citep[e.g.,][]{2009MNRAS.397.1153K}. 
This flux could in principle come from the precollapse stellar core. However such a progenitor is likely to experience significant internal magnetic torques \citep{2002A&A...381..923S}, causing it to rotate too slowly to be a viable GRB central engine \citep*{2005ApJ...626..350H}. Short-duration GRBs have similar jet powers; these events are associated with mergers of old neutron star binaries, which are unlikely to possess fields $\sim 10^{15}$~G before the merger.

Pre-existing large-scale magnetic flux also appears to be insufficient to drive jets in tidal disruption events (TDEs). In particular, the TDE candidate SW~1644+57 had a peak X-ray luminosity of $\simmore 10^{48}$ erg s$^{-1}$ \citep{2011Sci...333..199L}; after correcting for beaming, the true jet power $\Pj$ may be estimated as $\sim 10^{46}$~erg~s$^{-1}$. Then equation~(\ref{eq:Pj}) requires flux $\Phi\sim 10^{30} M_6 L_{\rm{j},46}^{1/2}$~G~cm$^2$, where $M_6$ is the black-hole mass in units of $10^6\, \Msun$.
This amount of flux is unlikely to come from the disrupted star \citep{2014MNRAS.437.2744T}.

In active galactic nuclei (AGN) and X-ray binaries, enough magnetic flux may be available at large radii. 
However, it is unclear if it can be advected onto the black hole.
For thin discs, inward flux transport requires the effective viscosity to be
much larger than the effective magnetic diffusivity, which is unlikely if both are due to turbulence driven by the magneto-rotational instability (MRI) \citep*{1994MNRAS.267..235L,1996ApJ...473..403H}; see \citet{2005ApJ...629..960S} and \citet{2008ApJ...677.1221R} for alternative scenarios. Direct flux advection may be possible if the disc is thick. However, the discs in AGN, even if geometrically thick close to the black hole, are expected to be very thin further out, as demonstrated by observations of water megamasers on the 0.1--1 pc scale \citep{1994ApJ...437L..99B}. Moreover, the most powerful jet sources, the flat-spectrum radio quasars, have large thermal disc luminosities $L_{\rm disc} \approx 0.1\, L_{\rm edd}$ \citep{1989ApJ...346...68S,2010MNRAS.402..497G}, implying radiatively efficient thin discs stretching down to small radii. 
In these objects the existence of large-scale flux near the black hole remains an open question.

Instead, here we propose that jets can be created by small-scale magnetic flux systems which are naturally amplified and sustained in a turbulent accretion disc; no large-scale flux is required. Magnetic field loops that connect the black hole to the inner disc can open up, resulting in black-hole-powered BZ jet episodes. Below we outline the jet formation process and present general-relativistic force-free electrodynamic simulations, demonstrating the magnetosphere's time evolution, including the effect of tearing-mode reconnection. \KP{Simulations of this process require a numerical scheme which can accurately evolve the magnetically dominated region outside the accretion disc; we therefore use the pseudospectral force-free code \textsc{phaedra} \citep*{2012MNRAS.423.1416P}, which has been extended to curved background spacetimes \citep{ParfreyPhDT}.}

\section{Jets from small-scale magnetic fields}
\label{sec:smallscale}

The accretion of plasma is thought to be driven by the MRI \citep{1991ApJ...376..214B}, which naturally amplifies magnetic fields until the magnetic pressure is a certain fraction of the gas pressure at the disc's midplane. The natural large length scale $l$ of the generated magnetic field is similar to the disc thickness $2 H$. We envision jet production as a three-stage process:
\begin{enumerate}
\item A magnetic field loop, of poloidal width $l$, buoyantly emerges from the disc surface \citep*{1979ApJ...229..318G}.
We consider a loop near the disc's inner edge, which is associated with the innermost stable circular orbit (ISCO). 
The footpoint of the loop closer to $\risco$ is carried with the disc material into 
the black hole, while the outer footpoint remains in the disc. The hole and disc are now coupled, and begin to transfer energy and angular momentum between them.

\item Differential rotation of the two footpoints twists the magnetic field lines, and pressure from the accumulated toroidal field forces the field lines to rise away from the hole and disc, `opening up' toward infinity. 
Energy and angular momentum are extracted from the hole, powering a relativistic outflow along the opening field lines.

\item On the accretion timescale $\tacc \sim l/\vr$, where $\vr$ is the radial velocity in the disc, the outer footpoint is eventually also advected onto the black hole with the incoming plasma. A current sheet forms along the centre line of the open loop, inevitably becoming unstable to reconnection. The jet power from the loop declines as the open field lines reconnect, driven by the continual accretion of magnetic flux. \KP{Reconnection will occur on a range of timescales, from $t_{\rm rec,min} \sim \rg/\epsilon c$, where the reconnection speed is $v_{\rm rec}=\epsilon c$, with $\epsilon\sim$ 0.01--0.1 \citep*[e.g.,][]{2010PhRvL.105w5002U}, to the time taken for accretion of the second half of the loop, $t_{\rm rec,max} \sim \tacc/2$.}
\end{enumerate}

A crucial consideration is whether the field lines coupling the black hole to the disc can remain closed or are forced to open. If they are closed no energy can magnetically be removed to infinity and no jet can be powered by the hole's rotational energy. \citet{2005ApJ...620..889U} has shown that for prograde Keplerian discs, even at high black-hole spin, these field lines can remain closed out to a certain radius $\rclose$ which is located beyond $\risco$. This sets a critical scale $\lcrit = \rclose - \risco \sim O\lp\rg\rp$. A jet will not be created unless the loop's width $l$ is greater than $\lcrit$, and will be destroyed once the loop's outermost footpoint is brought inside $\rclose$.
It is for this reason that jets are created more naturally by retrograde accretion flows: the differential angular velocity between the horizon and $\risco$ is so great that all coupled field lines are forced to open, removing the critical loop scale. 

We can now estimate the jet power and efficiency using standard relations for $\alpha$-discs \citep{1973A&A....24..337S}. The gas pressure at the disc midplane is given by $P = \rho \cs^2 = \rho \lp H/r \rp^2 \vk^2$, where the Keplerian speed $\vk=(\rg/r)^{1/2}c$.
The disc density $\rho$, half-thickness $H$, and accretion speed $\vr = \alpha \lp H/r \rp^2 \vk$ 
correspond to the mass accretion rate $\dot{M} = 4\pi r H \rho \vr$. The magnetic field in the disc is related to the gas pressure by $B^2 = 8\pi P/\beta = 2 \dot{M} \rg^{1/2}c/\alpha\beta H r^{3/2}$, and the poloidal magnetic flux in half an axisymmetric loop of size $l$ is $\Phi_l = 2\pi r (l/2) B_{\rm p}$, where $B_{\rm p}$ is the poloidal field component.
When this flux is brought onto the black hole, the instantaneous BZ power $L_0$ may be estimated from equation~(1). Setting
$r=\risco$, $a=1$, and $\rh = \rg$ gives
\beq
L_0 \approx  \Upsilon \frac{\Phi_l^2 a^2}{\rh^2} c \approx  2\pi^2\Upsilon \frac{l^2}{\alpha\beta H}\frac{\risco^{1/2}}{\rg^{3/2}}\left(\frac{B_{\rm p}}{B}\right)^2 \dot{M} c^2.
\label{eq:estimate}
\eeq
Using $\Upsilon = 10^{-3}$, $\alpha\beta = 1/2$ \citep{2000ApJ...534..398M,2009ApJ...694.1010G}, and $l = 2H$, we obtain the estimate
\begin{equation}
  L_0\approx 0.16\, \frac{H}{r} \left(\frac{\risco}{\rg}\right)^{3/2} 
                      \left(\frac{B_{\rm p}}{B}\right)^2 \dot{M} c^2.
\end{equation}
 This estimate suggests that accretion of small-scale magnetic loops $l\sim 2H$ 
 is capable of driving powerful jets, especially in the case of retrograde discs which 
 have $\risco = 9\, \rg$.

\section{Numerical simulations}
\label{sec:numerical}

\begin{figure*}
\begin{center}
\includegraphics[width=\textwidth, trim = 2.8mm 2.5mm 3mm 2.5mm, clip]{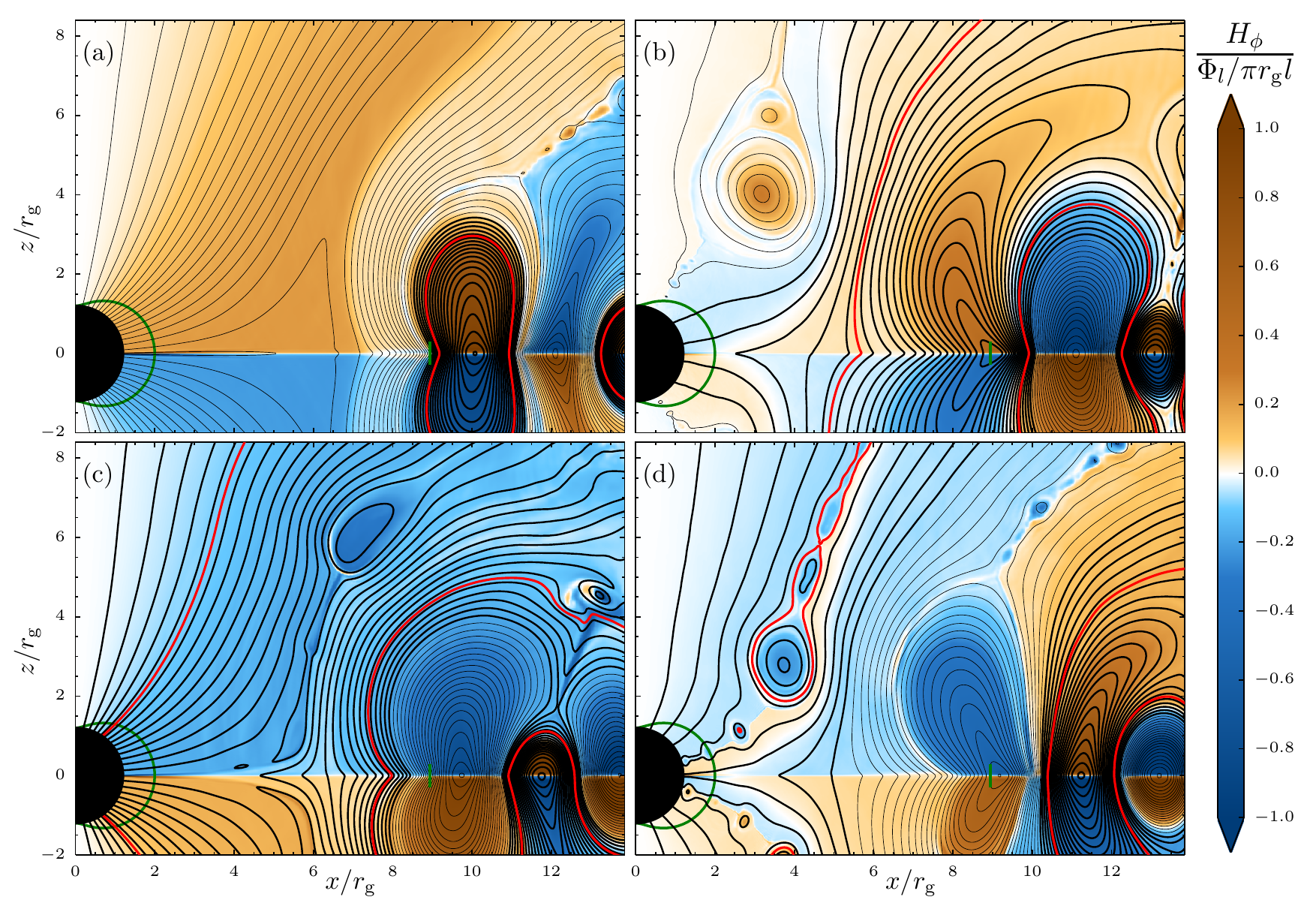}
\end{center}
\caption{\label{fig:fig1} Retrograde simulation. (a) $t = 592\rgc$. A fresh magnetic loop is poised just outside $\risco$. (b) $t = 794\rgc$. As the new loop begins to be accreted onto the black hole it compresses the preceding flux system, inducing reconnection. Poloidal field lines begin to expand outward. Peak jet power is reached at $t \approx 975 \rgc$. (c) $t = 1064\rgc$. Reconnection begins in the current sheet along the loop's centre line, ejecting plasmoids to infinity. (d) $t = 1170\rgc$. The reconnecting current sheet is brought onto the black hole. Thick (thin) black lines are poloidal field lines for initially clockwise (counter-clockwise) loops as viewed in the poloidal ($r$--$\theta$) plane, with one additional field line per clockwise loop highlighted in red; the green bar indicates the position of the ISCO and the green curve is the ergosphere boundary. $H_\phi$ is the poloidal current function, loosely equivalent to the toroidal magnetic field.}
\end{figure*}

We have investigated the above process with a set of time-dependent axisymmetric numerical simulations, performed in the force-free (high-magnetization) limit of MHD. The \textsc{phaedra} code has been augmented by several improvements, such as dynamic physical resistivity and more effective current-sheet capturing, which will be described in a future paper. Kerr-Schild coordinates are used, allowing the inner computational boundary to be placed inside the horizon, and therefore out of causal contact with the rest of the calculation. The black hole has spin parameter $a = 0.98$. The outer boundary is placed at $500\,\rg$ and an absorbing layer prevents reflection of outgoing waves. The grid resolution is $N_{r} \times N_\theta = 1024\times 512$, and the radial grid is strongly biased toward the black hole.

The initial conditions consist of a series of equal-flux magnetic loops, of width $l$ and alternating polarity, sourced from a thin disc in the equatorial plane; this configuration is a vacuum steady state. At $t = 0$ conducting massless plasma is injected everywhere (i.e., the force-free current term is activated) and an ideal heavy thin accretion disc begins to advect and twist the frozen-in magnetic field loops, \KP{with spatial velocity field $v^i = u^i/u^t = (v^r,0,\Omega_{\rm K})$, where $u^\alpha$ is the 4-velocity in Kerr-Schild coordinates and $\Omega_{\rm K} = \pm c/(r \sqrt{r/\rg}  \pm a \rg)$ is the Keplerian angular velocity for prograde ($+$) and retrograde ($-$) orbits}. Inside $\risco$ the gas, \KP{into which the magnetic field remains frozen}, plunges geodesically into the black hole. We describe here two simulations, one each for prograde and retrograde accretion flows. In both cases the accretion speed is \KP{a constant $v^r = -c/200$ outside $\risco$} and the loop width $l = 2\,\rg$.

In the prograde simulation, field lines are inflated by the Keplerian differential rotation while both footpoints are attached to the disc, but they collapse to a lower-energy closed state once their inner footpoints are advected onto the black hole. They no longer expand poloidally, but rather passively transfer energy and angular momentum from the black hole to the inner disc. The disc differential rotation drives large eruptive plasmoid ejections from the next-to-innermost flux system. \KP{There is no outflow from the black hole, because the loop scale is smaller than the critical scale, $l < \lcrit$; in this case we find $\lcrit \approx 3.2 \,\rg$.}

\begin{figure}
\begin{center}
\includegraphics[width=3.35in, trim = 3mm 2.5mm 2.5mm 2.5mm, clip]{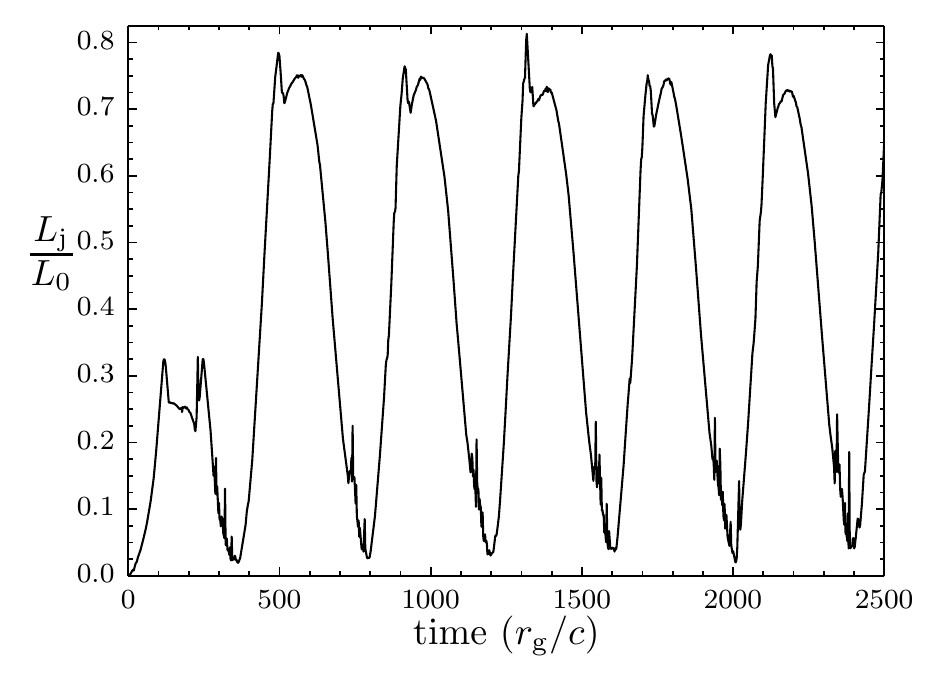}
\end{center}
\caption{\label{fig:fig2} Jet power in units of the fiducial power $L_0 = \Upsilon \Phi_l^2 a^2 c/\rh^2$ over five full jet-launching cycles.}
\end{figure}

The retrograde configuration behaves entirely differently. Consider the evolution from when a flux loop first reaches the ISCO (Fig.~\ref{fig:fig1}a). As the loop's leading edge is accreted through the ISCO the entire flux structure begins to open up, due to the fast differential rotation between footpoints near the black hole and those still rooted in the disc. 
The loop begins to be dragged onto the black hole, pushing together oppositely directed field lines ahead of it and inducing magnetic reconnection (Fig.~\ref{fig:fig1}b). At this time the absolute flux through the horizon is at a minimum and therefore so is the jet power (Fig.~\ref{fig:fig2}). 
The differential rotation across a field line changes sign as its leading footpoint approaches the black hole and experiences the frame dragging effect in the direction opposite to the disc's retrograde Keplerian rotation. As a result the toroidal magnetic field in the loop changes sign.
As more of the loop is brought onto the hole the flux through the horizon increases and the integrated outgoing Poynting power climbs steadily. The angular velocity difference between the black hole and the retrograde Keplerian disc is so great that, despite the ability of the magnetic field to slip through the horizon, all field lines threading the hole are forced to open toward infinity (Fig.~\ref{fig:fig1}c).
By the time of peak jet-power output almost all of the available magnetic flux has opened and the field configuration is similar to Fig.~\ref{fig:fig1}a (with loop polarities reversed).

A current sheet is formed through the centre of the loop when the poloidal field inflates and becomes approximately radial. The current sheet becomes unstable to reconnection via the tearing mode, causing magnetic energy to be dissipated into radiation and particle kinetic energy (heating and acceleration). We observe that reconnection begins near the intersection of the current sheet and the equatorial disc, at $\sim4\,\rg$. Initially the current sheet intersects the disc and all of the plasmoids created by the tearing process are ejected from the system (Fig.~\ref{fig:fig1}c). Eventually the current sheet base is carried onto the black hole by the plunging plasma; the sheet now extends through the horizon, and some of the plasmoids are swallowed by the hole (Fig.~\ref{fig:fig1}d). Soon all of the loop's field lines have both footpoints on the horizon, and the reconnection continues to be driven by the Maxwell pressure of the next flux system being accreted, until all of the loop's magnetic energy has passed in through the horizon or been expelled, and the next loop has replaced it on the black hole. \KP{The entire reconnection process takes about $\Delta t \approx 185 \rgc \approx l/2\vr = \tacc/2$}. Large plasmoids are sometimes formed by the coalescence of smaller plasmoids, especially in the later stages of reconnection (Fig.~\ref{fig:fig1}b). Although most of the dissipation occurs inside $\risco$, reconnection also takes place beyond this point, as coronal flux systems entirely frozen into the disc expand under the Keplerian shear and form current sheets (Fig.~\ref{fig:fig1}, all panels).

The anticipated maximum jet power is $L_0=\Upsilon \Phi_l^2 a^2 c/\rh^2$ (Section~2). To estimate the BZ power $\Pj$ in our simulation we consider the energy flux through the black hole's surface. The total flux of energy-at-infinity through the horizon is $L = L_{\rm out} - L_{\rm in}$, where $L_{\rm out}$ ($L_{\rm in}$) is the total luminosity of outgoing (ingoing) energy. We associate the jet power $\Lj = L_{\rm out}$; it is plotted in units of $L_0$ in Fig.~\ref{fig:fig2}. 

\KP{At $t=0$ much of the first loop's flux lies inside the plunging region and cannot inflate fully before the onset of reconnection; hence the first cycle ($t = 0$--$350\rgc$) has a reduced peak power. For all other cycles the power reaches a maximum of $\approx 0.75 \,L_0$.}
Averaging over five complete cycles  we find the jet power to be $\langle\Lj\rangle = 0.43 \,L_0$. \KP{The outflowing electromagnetic energy survives to large radii; e.g., the average power through a sphere of radius $r = 100 \,\rg$ is $0.51\, L_0$, where the increase in power is supplied by the disc.}

\vspace{-0.4cm}

\section{Discussion and Conclusions}
\label{sec:discuss}

We have argued that small-scale magnetic fields, sourced in an accretion disc where they are amplified by MRI turbulence, can launch powerful relativistic jets when coupled to a rapidly rotating black hole. 
In this scenario the existence and transport of large-scale net flux are unnecessary. For prograde discs, there is a minimum poloidal length scale $\lcrit$ below which the magnetic loops can remain closed when connecting the black hole to the disc, preventing jet production. There is no such minimum loop scale for retrograde accretion, allowing these flows to power jets even for lower disc thicknesses (if $l \propto H$). Retrograde discs also naturally contain loops with larger flux $\Phi_l$ near the disc's inner edge, and so can produce more powerful jets since $\langle\Lj\rangle \propto \Phi_l^2$.

Prograde discs can, however, create jets via this mechanism provided field loops of sufficient size are present. 
Several effects are expected to mitigate in favour of prograde jet launching. 
A coronal inverse-cascade would increase the maximum loop scale \citep{2008ApJ...682..608U}, potentially beyond the critical value. Inflated hole-disc coupling field lines may diffuse outward due to strong magnetic tension forces; the further they diffuse the greater the differential rotation between the line's footpoints, possibly resulting in diffusion to beyond $\rclose$ if the accretion velocity is sufficiently low.

Retrograde accretion has previously been invoked to explain aspects of the radio-loud quasar population \citep*{2009ApJ...699..400G}. Retrograde accretion should occur naturally in a fraction of TDEs, AGN, and wind-fed X-ray binaries. While determining observationally the direction of an accretion flow (closely related to measuring the spin of the black hole) remains difficult, there are suggestions of retrograde accretion in a small number of radio galaxies \citep{2009ApJ...700.1473S,2011ApJ...734..105S} and microquasars \citep{2013ApJ...778..155R,2014MNRAS.439.1740M}.

The ability of prograde accretion flows to maintain closed black hole-disc field lines may explain the absence of jets in known high-spin prograde accretion systems; e.g., iron line measurements of several Seyfert galaxies indicate rapidly rotating black holes ($a > 0.8$) and prograde accretion \citep{2011ApJ...736..103B,2013Natur.494..449R}. 
This jet-quenching mechanism may also operate in those black hole X-ray binaries which have no jet in the soft state, when a thermal disc is observed down to a few $\rg$ \citep[e.g.,][]{2014MNRAS.442.1767P}.
A jet may form in the hard state because the inner accretion flow is geometrically thick, supporting flux loops larger than $\lcrit$, or because large loops are created by an inverse-cascade in a magnetically active X-ray corona.

The jets produced by the proposed mechanism are naturally highly variable, on both the loop accretion and the reconnection timescales. Reconnection occurs constantly above the disc as flux systems are energized by the azimuthal shear, expand, and form current sheets. However most of the dissipation takes place very near the black hole and along the axis, when greatly inflated field loops, storing large magnetic free energy, suffer reconnection of almost all their field lines; this is true for both prograde and retrograde configurations. The reconnection converts magnetic energy into radiation, plasma thermal energy, and kinetic energy of accelerated particles, and may be the ultimate source of X-ray coronae in AGN and X-ray binaries. Our model naturally places the hard X-ray source
at the base of an outflow. An outflowing corona was previously proposed to explain 
the hard state spectrum of accreting black holes \citep{Beloborodov:1999aa}.
The concentration of the dissipation at small radii is consistent with the small X-ray source sizes, $\lesssim 10\,\rg$, deduced from quasar microlensing \citep{2008ApJ...689..755M}. Reconnection at larger distances, between the varying-polarity flux ejected from the accretion of different loops, may be responsible for blazars' gamma-ray emission \citep{2013MNRAS.431..355G}.

Our analysis has been two-dimensional---in reality several flux systems, of azimuthal extent $\Delta\phi \simmore H/r$, will interact with the black hole concurrently, each contributing to the overall large-scale jet and peaking independently in power output. These loops will interact with one another, especially near the black hole where they are pressed together by accretion and significant flux rearrangment due to reconnection and interchange instability becomes likely; we intend to study these effects with three-dimensional simulations. Finally, in our simulations the velocity field of the disc plasma has been imposed by hand. This kinematic approximation is likely to break down near the black hole, where the compressed magnetic field is strong. Especially in our most efficient retrograde scenario, the magnetic field near the hole may become strong enough to impede accretion \citep[e.g.,][]{1986bhmp.book.....T}; here again the interaction of azimuthally separated flux systems arriving on the hole at different times will become important.

KP is supported by the Max-Planck/Princeton Center for Plasma Physics. DG acknowledges support from NASA grant NNX13AP13G. AMB acknowledges support from NASA grant NNX13AI34G. Simulations were performed on the computational resources supported by the PICSciE TIGRESS High Performance Computing Center.

\bibliographystyle{mn2e}

\end{document}